\documentclass[twoside]{article}
\usepackage{fleqn,espcrc2,epsfig,rotating}

\title{The Second Moment of the Pion Light Cone Wave Function}

\author{
Luigi Del Debbio\address[PISA]{Dipartimento di Fisica, Universit\'a di Pisa, P.zza Torricelli 2, Pisa, Italy},
Massimo Di Pierro\address[FNAL]{Fermilab, PO Box 500, Batavia, IL 60563, USA}
	\thanks{Talk presented by Massimo Di Pierro} and
Alex Dougall\address[LIV]{Theoretical Physics, Dept. of Mathematical Sciences, Univ. of Liverpool, Liverpool, L69 3BX, UK}}

\begin{document}

\newcommand{\text}[1]{\textrm{\footnotesize #1}}

\begin{abstract}
We present a preliminary result for second moment of the light cone wave 
function of the pion. This parameter is the subject of a discrepancy 
between theoretical predictions (coming from lattice and sum rules) 
and a recent experimental result (that remarkably agrees with purely 
perturbative predictions). In this work 
we exploit lattice hypercubic symmetries to remove power divergences and, 
moreover, implement a full 1-loop matching for all the contributing operators.
\vspace{1pc}
\end{abstract}

\maketitle

\section{INTRODUCTION}

The light cone wave function of the pion, $\phi(x,Q^2)$ is the 
probability amplitude of finding a parton of a pion 
(moving with momentum $\mathbf{p}$ in a light cone frame) with parallel 
momentum equal to $x\mathbf{p}$ and transverse momentum less than 
$Q$. 
This wave function incorporates non-perturbative physics and plays 
an important role in exclusive hard scattering processes and in 
non-leptonic decays of heavy mesons.
One example of application is the electromagnetic form factor 
of the pion, $F(q^2)$, defined by
\begin{equation}
\langle \pi(\mathbf{p}') | \bar q \gamma_\mu q | \pi(\mathbf{p}) \rangle =
F((p-p')^2)(p+p')_\mu
\end{equation}
This form factor can, in fact, 
be written in terms of $\phi$ and $T_H$ (the perturbative 
scattering amplitude for the constituents) as
\begin{equation}
F(Q^2)\!= \!\!\int \! \phi^\dagger(x,Q^2)T_H(x,y,Q^2)\phi(y,Q^2)
\textrm{d}x\textrm{d}y
\end{equation}

More formally the light cone wave function can be defined as~\cite{chernyak}\cite{brodsky}
\begin{equation}
\phi^{ab}_{\alpha\beta}(x,Q^2) = F.T. \langle 0 |
T\{q^a_\alpha(z_1), \bar q^b_\beta(z_2)\} | \pi \rangle
\end{equation}
where F.T. indicates a Fourier transform on $z_1$ and $z_2$ assuming 
\begin{itemize}
\item the sum of the parallel components of the momenta of the two partons 
is equal to the total pion momentum $\mathbf{p}$.
\item the transverse components have been integrated out up to momentum $Q$.
\end{itemize}

The $n$-th moment of $\phi$ is defined as
\begin{equation}
\langle \xi^n \rangle  = \int_0^1 \xi^n \phi(\xi, Q^2)\textrm{d}\xi
\end{equation}

In a typical lattice determination of such quantity the cut-off 
is provided by the lattice spacing, $Q \simeq a^{-1}$.

We finally wish to remark that $\langle \xi^0 \rangle=1$ is fixed by
a normalization condition and $\langle \xi^1 \rangle=0$ because the wavefunction
is symmetric under G-parity. Therefore $\langle \xi^2 \rangle$ is the
first non trivial moment of $\phi$.

This second moment can be related to
\begin{equation}
\langle 0 |O_{\mu \nu \rho}| \pi(\mathbf{p}) \rangle = 
f_\pi \langle \xi^2 \rangle p_\mu p_\nu p_\rho + ...
\label{definition}
\end{equation}
where the ellipsis indicates divergent terms,
\begin{equation}
O_{\mu \nu \rho} = \bar q \gamma_\mu 
\gamma_5 \stackrel{\leftrightarrow}{D}_\nu \stackrel{\leftrightarrow}{D}_\rho q'
\end{equation}
and
\begin{equation}
q\stackrel{\leftrightarrow}{D}q' \stackrel{def}{=}q\stackrel{\rightarrow}{D}q'+ q\stackrel{\leftarrow}{D}q'
\end{equation}

\section{COMPUTATION}

Our computation was carried on 154 quenched gauge 
configurations generated by the UKQCD collaboration 
using the Wilson gauge action. We use the 
Clover $O(a^2)$ improved action for the light quarks.

The parameters of our computation are:
\begin{itemize}

\item Volume equal to $24^3\times 48$.

\item $\beta=6.2$ which corresponds to an inverse lattice spacing of
$a^{-1}=2.67\pm0.10$GeV.

\item $\kappa$ values 0.13460, 0.13510, 0.13530 (corresponding
to pseudoscalar masses of 748, 574 and 490MeV respectively)

\item $\kappa_{crit}=0.13582$ and $c_{SW}$=1.61

\end{itemize}

For technical reasons we choose not to improve the operators and do not 
smear the light quarks. The latter choice is motivated by the fact 
that the local axial current seems to have better superposition with the 
pion than other smeared operators we tried.

In order to calculate any moment of the pion light-cone wavefunction,
we are required to  study the matrix elements of lowest twist local 
operators between pion and vacuum. In a continuum world 
these operators are classified by their representation under the 
group of Lorentz, parity and charge conjugation. 
In the lattice-discretized world the Lorentz group is broken to 
$\mathcal{H}_4 \in O(4)$, the hypercubic group. Hence
lowest twist local operators can mix with higher dimensional operators and 
introduce power divergences. 

We choose a particular combination of the lattice operators that:
\begin{itemize}
\item transforms under an irreducible representation of the hypercubic group
\item does not mix with higher dimensional operators.
\end{itemize}
In particular we choose $O_{\mu[\nu,\rho]}$ with $\mu\neq\nu\neq\rho\neq\mu$.
which transforms as $\bar{(\frac12, \frac12)} \otimes \mathbf{8}^{+}$ under 
$\mathcal{H}_4$. The $\mathbf{8}^{+}$ irreducible representation would naively 
generate a term proportional to 
\begin{equation}
p_\mu \left[\frac{p_\nu^2+p_\rho^2}{2} - (\epsilon^{\mu\nu\rho\sigma} p_\sigma)^2 \right]
\label{mixing}
\end{equation}
This contribution vanishes for our matrix elements because of its parity. 
We introduce the following definition:
\begin{equation}
R = \frac{C_2^O (t,{\mathbf{p}})}{p_1 p_2 C_2^A(t,{\mathbf{p}})}_{\mathbf{p}=(1,1,0)} 
\label{R1}
\end{equation}
where
$C_2^Q(t,{\mathbf{p}})$ is the spatial Fourier transform 
(at momentum $\mathbf{p}$) of $\langle Q(x) \bar q \gamma_5 \psi q' (0) \rangle$, 
$O$ is a short hand 
notation for $O_{\mu[\nu,\rho]}$ and $A=\bar q \gamma_5 q'$ is the usual 
axial current.
For large $t=x_0 \rightarrow \infty$, $C_2(t)$ asymptotically approaches 
\begin{equation}
C_2 \simeq \frac{Z_A}{2 E(\mathbf{p})} \langle 0 | Q(0) | \pi(\mathbf{p}) \rangle e^{- E(\mathbf{p}) t }
\label{R2}
\end{equation}
Using eq.~(\ref{definition}),  eq.~(\ref{R1}) reduces to
\begin{equation}
R\simeq \langle \xi^2 \rangle^{lattice} = 
\frac{Z^A}{Z^O}\langle \xi^2 \rangle^{\overline{\text{MS}}}.
\end{equation}
the desired second moment (without divergences) times a corrective matching 
factor to be determined perturbatively.

\begin{figure}
\begin{center}
\begin{turn}{270}
\epsfig{file=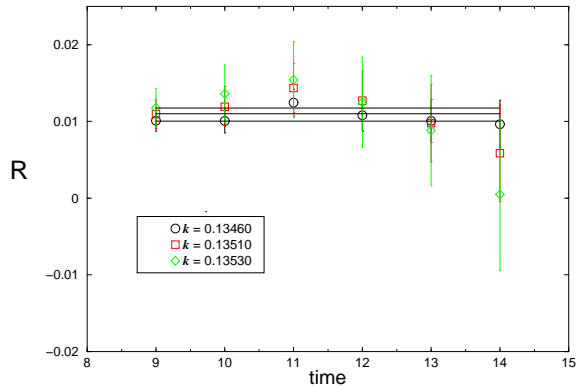, height=75mm}
\end{turn}
\end{center}              
\caption{The Plot shows R as function of $t$, the timeslice, and $\kappa$. 
The best fits are included (for each value of $\kappa$).\label{fig1}}
\end{figure}

\section{MATCHING}

A big part of this work was the determination of the matching factor $Z^O$ 
for $O=O_{\mu \nu \rho}$. The computation was performed assuming a gluon mass 
as regulator and on-shell external quarks with non-zero momentum. All 41 
relevant diagrams were expanded in Taylor up to second order in $\mathbf{p}$,
compatibly with eq.~(\ref{R1}).

As result of our computation we find that
\begin{equation}
\frac{Z^O}{Z^A}=1.518
\end{equation}
and the coefficients are evaluated at $\alpha_s(q^*)$ for $q^*=2/a$.
We tried varying $q^*$ by factors of two and this gives us an estimate of
the error in the matching of the order of 10\% that we include in our final 
result.

\section{RESULTS}

The result of our calculation is 
\begin{eqnarray}
\langle \xi^2 \rangle ^{lattice}_{Q=2.67GeV} &=& 0.185 \pm 0.032 \\
\langle \xi^2 \rangle ^{\overline{\text{MS}}}_{Q=2.67GeV} &=& 0.280 \pm 0.049 ^{+0.030}_{-0.013}
\end{eqnarray}
(the first error is statistical and the second is systematic due to matching, quenching error is not included).
These numbers should be compared with results from sum rules
\begin{eqnarray}
\langle \xi^2 \rangle ^{\overline{\text{MS}}}_{Q=5GeV\phantom{2.}} &=& 0.40 \pm 0.05
\end{eqnarray}
with preceding independent lattice results
\begin{eqnarray}
\langle \xi^2 \rangle ^{lattice}_{Q \simeq 1GeV\phantom{4.}} &=& 1.37 \pm 0.20 ~\cite{andreas} \\
\langle \xi^2 \rangle ^{lattice}_{Q \simeq 1GeV\phantom{4.}} &=& 0.25 \pm 0.10 ~\cite{sachrajda} \\
\langle \xi^2 \rangle ^{lattice}_{Q=2.4GeV} &=& 0.10 \pm 0.12 ~\cite{gupta} 
\end{eqnarray}
and the exact asymptotic value
\begin{eqnarray}
\langle \xi^2 \rangle_{Q=\infty\phantom{.GeV}} &=& 0.2
\end{eqnarray}
(confirmed by the Fermilab experiment E791, performed at $Q \simeq 3-4$GeV).

We find that our value for the second moment is smaller than 
sum rule predictions and is closer to the asymptotic value.

On the one side we conclude that the present computation has better 
control of statistical and perturbative errors than previous computations.
On the other side we strongly feel that dynamical quarks may be playing an
important role in this process and quenching errors are yet to be estimated 
and removed. We also believe that the same analysis should be repeated 
for different values of the lattice spacing in order to study the $Q^2$ 
dependence of $\langle \xi^2 \rangle$.

\section*{ACKNOWLEDGMENTS}
We all wish to thank UKQCD for making the gauge configurations available to 
us and for giving us access to the Cray T3E (where part of the computation
was performed).
We also wish to thank Chris Sachrajda for his invaluable contribution and
the University of Southampton where this research 
started. Those of us who no longer work in Southampton thanks their present
host institutions for support and funding in continuing this research.

\end{document}